  \documentstyle [12pt,twoside]{article}
  
\begin{document}
\begin{flushright}
UBCTP--93--005
\end{flushright}
\vspace*{0.5cm}
\centerline{\large{\bf Gravity Wave Watching}}
\vspace*{1cm}
\centerline{\bf Redouane Fakir}
\vspace*{0.5cm}
\centerline{\em Moroccan Center for Scientific Research}
\centerline{\em 52 Ave. Omar Ibn Khattab}
\centerline{\em BP 1364 Agdal, Rabat}
\centerline{\em  MOROCCO}
\vspace*{0.5cm}

\centerline{\em Cosmology Group, Department of Physics}
\centerline{\em University of British Columbia}
\centerline{\em 6224 Agriculture Road}
\centerline{\em Vancouver, B. C. V6T 1Z1}
\centerline{\em CANADA}
\vspace*{1.cm}

\centerline{\bf Abstract}
\vspace*{0.5cm}
It is suggested that gravity waves could, in several cases,
be detected by means of already (or shortly to be) available
technology, independently of current efforts of detection.
The present is a follow-up on a recently suggested detection
strategy based on gravity-wave-induced deviations of null
geodesics. The new development is that a way was found to
probe the waves close to the source, where they are several
orders of magnitude larger than on the Earth.
The effect translates into apparent shifts in stellar angular
positions that could be as high as $10^{-7}$ arcsec, which is
just about the present theoretical limit of detectability.
\clearpage

\centerline{\bf I. Introduction}
\vspace*{0.5cm}
It was argued recently that gravitational waves could,
{\em in principal,} be detected through a new
non-mechanical effect [1,2]. This effect can be easily
understood by replacing the usual textbook case of
gravity-wave-induced  variations in proper distances
between initially static massive particles, by the
analogous case for massless particles. One is then
monitoring deviations in null, rather than in timelike
geodesics. When the massless particles are photons,
the timelike geodesics can be light rays coming from
distant (e.g. stellar) sources, or perhaps from
man-made space-based laser sources. Thus, what is
to be measured in this approach, are gravity-wave-induced
shifts in {\em apparent} angular separations of appropriate
light sources. This effect should not be confused with
the much smaller shifts in {\em real} angular
separations of light sources, due to gravity waves
actually hitting the light source \cite{kn:brgr}. In \cite{kn:fakir1},
it is
the {\em photons} from the source (not the source itself)
that encounter the waves,
an that could happen
arbitrarily close to the Earth. Nor should the effect be confused
with yet another smaller shift, that
due to the overall focusing of geodesics by the energy of
the gravity waves.

For an introduction to the physics of gravity waves
and to the many proposed, and the few fully developed detection
strategies, the reader is referred, in
[4-22], to some reviews and key articles in the field.

Consider, now, the following illustration of the
effect, as discussed in \cite{kn:fakir1}:
A gravitational pulse ``with memory of position" [3,19,20,23-27] hits
the Earth, coming from a supernova explosion, say, in the
Northern sky. Alerted by the electromagnetic flash, the
observer is now aware that a
gravity wave has just whipped passed the Earth, and is
traveling towards the Southern sky. For many years to
come, light reaching the Earth from most sources in that
hemisphere will have {\em been through the gravity wave}.
Thus, if arbitrarily small angular shifts were detectable,
the receding gravitational pulse could be ``followed'' for
an extended duration. For most other schemes of gravity-wave
detection, the same pulse would leave a single blip on the
detector's record, and then would be lost for ever.
Here, the gravitational pulse,
all along its receding journey, keeps interacting
with photons that eventually reach the observer.
Hence, in a sense, the photons act as detectors
that interact with the pulse
farther and farther away from the Earth.

At this point, the reader could be wondering what all the
fuss is about, and she (or he) would be right, as far as this
particular illustration is concerned. For the waves get
only dimmer (though at a negligible rate,) as they keep
moving away from us, and the expected apparent angular
shifts (of the order of the wave's strength at the
location of the photon-graviton
interaction \cite{kn:fakir1},) are over ten orders of magnitude smaller than
what present day angular resolution astrometry can handle.

However, as we shall now see, exploiting the same physics
as above, but in a sense inverting the argument, one can
arrive at a scheme for detecting gravity waves directly,
using present-day resolution-power technology.

Consider the following configuration. A steady
source of gravity waves (e.g. a neutron star or a binary
star, call it for short SG,) is giving off radiation with each
polarization component having the form
\newline\begin{equation}
h = {H\over r}\exp \{i\Omega (r-t)\} \ \ ,
\end{equation}\newline
where r is the distance to the center of the gravity-wave
source SG, and $\Omega$ is of the order of the time frequency
of that source. (E.g. $\Omega =2\omega$ when the source is a
system with rotation frequency $\omega$.)
 $H$ is a constant that can be roughly estimated
in the quadrupole approximation by
\newline\begin{equation}
H \sim MR^{2}\Omega^{2} \ \  ,
\end{equation}\newline
where $M$ and $R$ are the mass and size of SG.

Now, consider a light source (dubbed SL hereafter,)
situated even further away
from us, along the line Earth-SG. (A situation where such a good
alignment might happen naturally will be discussed below.)
The photons we receive
on the Earth from SL have, at some point in the past,
crossed an area of relatively strong gravity waves.
The maximum amplitude experienced by a photon is
$h_{max} \sim H/b$, where $b$ is the impact parameter
with respect to the center of SG. From the conclusions
of \cite{kn:fakir1}, there is then  some reason to suspect that the
position of SL, as observed from the Earth, is
continuously experiencing apparent angular shifts with
an amplitude proportional to $h(r=b)$. A priori, this effect could
be larger than the one in \cite{kn:fakir1} by as many orders of
magnitude as there is in the ratio $r_{E}/b$, where $r_{E}$
is the distance from SG to the Earth.

 Upon inspection, it
turns out that rather different considerations apply here
than for the plane wave with memory of \cite{kn:fakir1}, leading to a
modified formula for the expected angular shifts. However, the calculations
below show 1) that the effect here also is physical, 2)
that its magnitude, in not too unlikely optimal situations,
falls close to present limits of detectability.

Before proceding to the next illustration, note that the angular
shifts studied here are variations not about the actual
position of SL, but about that position as displaced by
 non-radiative light deflections.
However, those non-radiative deflections are either
virtually constant in time (as in the example above,) and are
therefore irrelevant here (besides being actually unmeasurable,)
or, as in the illustration below, they have a well-known time
dependence and can be easily subtracted. In any case,
 Schwarzschild angular shifts, for most interesting
candidate sites, have a much larger
time-scale than (and therefore should not interfere with potential
measurements of) the expected shifts from gravity waves.

Now, let our gravity-wave source SG be a rotating neutron star.
A good candidate should be rotating rapidly enough to yield waves
with a sufficient strength (see (2)), but not so rapid that the light
crossing its gravity waves would shift too quickly for the
sensitivity limit of the observer's hardware. (In the
case of insufficient integration time, the overall effect
of the waves would be a blurring of SL's image, with a caracteristic
signature.)

As for the role of the light source SL, let be played here by a
companion star, locked with our neutron star SG into a binary
system. Different SL candidates (ordinary main-sequence stars,
 pulsars, etc.,) will have different observational
merits. Among the factors that come into play are the apparent
magnitude of SL  (shorter minimal integration times for larger
magnitudes,)
and its wavelength (better angular resolution for longer waves,
best resolution for radio waves.) Because quite different
considerations apply for each type of companion SL,
 we limit ourselves here to deriving the main equations
describing the effect,
and obtaining orders of magnitude for reasonable optimal situations.
A more detailed study of some promising actual astronomical sites,
will be
published elsewhere \cite{kn:fakir4}.

Taking
a rapid pulsar as the companion SL (to the slower neutron star SG,)
offers a particularly good opportunity for following the time
evolution of the various parameters involved: Since it can be
known to an excellent precision where each component of the
binary pulsar is situated at any given moment (time resolution much
finer than the period of either the neutron star SG or the pulsar SL,)
one has an excellent time axis over which to draw eventual
light deflection observations (see below.)

Just such a system has
been considered recently to try and detect gravity waves through
an effect that has little to do with the angular shifts studied here
\cite{kn:fakir3}.
In that paper, it is the Shapiro
time-delay effect of the neutron star's gravity waves on the
pulsar's frequency that is considered. It is found to fall not too
far from
present limits of time-resolution power. In fact, the main problem
there is less likely to be the time resolution than the photometric
sensitivity, most pulsars being very dim radio sources (some notable
exceptions are considered in [28].)

Finally, we should perhaps call the attention
on one more possible source of confusion. In the case (referred to as
case-two hereafter,) where the
gravity-wave source SG is a neutron star and the light source
SL an ordinary star or perhaps a pulsar, there are two quite
different types of gravity waves involved. One is the
neutron star SG's radiation, which has a frequency of
typically several or many Hertz. The other is the gravitational
radiation generated by the binary system
as a whole. These much longer waves (minimal period of a few hours,)
are the ones
probed in the case (case-one,)  where the binary system
as a whole
 was playing the role of SG, the light source SL being some much
farther, unrelated body. In the following, when SG and SL are the
two components of a binary (case-two,) the lower-frequency radiation of
the whole binary system will not be considered. When we speak
of the near-zone (distances less than one gravitational wavelength,)
we mean the near-zone of SG, which is the neutron star in case-two,
and the whole binary in case-one.

 This is not to say that it
would not be interesting to consider the radiation of the whole
binary even, in case-two. But, because we would then be
not only in the near-zone of the source, but within the source itself,
we would have to address the issue of disentangling dynamic Newtonian
 from relativistic
contributions in that case [28].
\clearpage
\centerline{\bf II. Equations for the effect}
\vspace*{0.5cm}

Our first simplification will be to consider the effect of only
one polarization component of the gravity waves, namely the one
that distorts
geodesics in directions that are parallel to the plane of rotation of
the gravity-wave source SG (that is the equatorial plane when SG is
a neutron star, and the orbital plane when it is a binary.)
This simplification does not affect substantially our conclusions,
 since the other
component of the waves (the one with effects that are parallel to the
rotation axis,) is usually of the same order of magnitude and frequency.
(Although, attention to this other component can be extremely useful
in disentangling radiative from dipole Newtonian effects in the near
zone of SG.) Thus, we will be studying movements completely contained
in the source's rotation plane.

With these simplifications, the line element, in the spherical
transverse-traceless gauge, has the simple form:
\newline\begin{equation}
dt^{2} - dr^{2} - gd\phi^{2} = 0 \  \ ,
\end{equation}\newline
where
\newline\begin{equation}
g\equiv (1+h)r^{2} \  \ .
\end{equation}\newline
Here the coordinates are centered on SG and light rays stretch from
$\phi\sim 0$ to $\phi\sim \pi$. $h$ is the gravitational-wave strength
given by (1). As mentioned above, the Schwazschild component of
the metric
is not included, its contribution being either irrelevant (e.g. when
SG and SL are very distant unrelated objects,) or easy to subtract
from the observations (e.g when SG and SL are locked into a binary
system.) For the flat  ($h=0$) spacetime, of course, our equations
should describe straight trajectories stretching from $\phi =0$ to $\phi
=\pi$, with closest encounter with SG at $r=b$, where $b$ is
the ``impact parameter''.

One can either stick to these  polar coordinates or, alternatively,
introduce a null coordinate $v\equiv r-t$. It turns out that the
gain  from such a change for this particular problem is not
considerable. So we shall stick here with $t$, $r$ and $\phi$.

Null geodesics are described by the equations
\newline\begin{equation}
{dp^{t}\over d\lambda} + {1\over 2} g_{,t}p^{\phi \ 2} = 0 \  \ ,
\end{equation}\newline
\begin{equation}
{dp^{r}\over d\lambda} - {1\over 2} g_{,r}p^{\phi \ 2} = 0 \  \ ,
\end{equation}\newline
\begin{equation}
{dp^{\phi}\over d\lambda} + {g_{,t}\over g} p^{t}p^{\phi}
+ {g_{,r}\over g} p^{r}p^{\phi} - {1\over 2} {g_{,\phi}\over g}
 p^{\phi \ 2} = 0 \  \ .
\end{equation}\newline
Here $p^{\alpha}\equiv dx^{\alpha} /d\lambda $ are the photon's
momenta. Commas indicate ordinary partial derivatives.

Our aim is to determine $\phi$ as a function of $r$. We start
by calculating $p^{\phi}$. We see immediately  from (7) that,
even for a finite $h$, there are straight line solutions,
namely $p^{\phi} = 0$. These are actually trajectories made of two
(one incoming and one outcoming) components, with a discontinuous
kink at $r=0$. This is telling us that here, as in the plane wave
case \cite{kn:fakir1}, photons that hit the wave fronts at right
 angles suffer
no radiative deflection.

For $p^{\phi}\neq 0$, that is, for non-vanishing impact parameters,
one gets from (7) the simple expression
\newline\begin{equation}
p^{\phi} = {L\over g} \  \ ,
\end{equation}\newline
where $L$ is an integration constant.

Let us write (6) in the form
\newline\begin{equation}
p^{r}={1\over 2}
\int^{\phi}_{C} g_{,r} p^{\phi} d\phi \  \  ,
\end{equation}\newline
where $C$ is a constant to be determined shortly.
Dividing this equation by (8) and using (1,4),
one obtains
\newline\begin{equation}
{dr\over d\phi} = {p^{r}\over p^{\phi}}
 = (1+h) r^{2}  \int^{\phi}_{C} \
 {2 + (1+i\Omega r)h\over 2(1+h)r}  \ d\phi \  \  .
\end{equation}\newline

In the absence of gravity waves, one should be able to recover
the straight-line solution
\newline\begin{equation}
r_{flat} = {b\over \sin\phi} \  \  .
\end{equation}\newline
This fixes the value of $C$ at $\pi/2 \ $,
as one can see immediately from (10,11).

We now switch to the radial variable $u\equiv 1/r$.
Then, deriving (10) with respect to $\phi$, and retaining
only terms that are linear in $h$, we find
\newline\begin{equation}
u'' + u = -h' \int^{\phi}_{\pi/2} u d\phi
-{h\over 2} (u+i\Omega) \  \ ,
\end{equation}\newline
where primes indicate derivatives with respect to $\phi$.

In accordance with our linear treatment,
the fluctuation $h$ can be expressed in terms
of $\phi$ only  by substituting, in (1), the expressions
for $r$ and $t$ corresponding to the vacuum straight-line
solution. These expressions are (11) and
\newline\begin{equation}
t_{flat} = -b \  {\cos\phi\over\sin\phi} \  \ .
\end{equation}\newline
(The origins of coordinates in (11,13) are chosen in such
a way that $t_{flat}=0$ and $r_{flat}=b$ at $\phi=\pi/2$.)
Thus, we have
\newline\begin{equation}
h = {H\over b} \sin\phi \
\exp\left\{ i\Omega b \  {1+\cos\phi\over\sin\phi} \
\right\} \  \ ,
\end{equation}\newline
and
\newline\begin{equation}
h' \equiv {dh\over d\phi}
= {h\over\sin\phi} \left(
\cos\phi - i\Omega b \ {1+\cos\phi\over\sin\phi} \right) \  \ .
\end{equation}\newline
On the other hand, the integral in (12) is only needed to
zeroth order. Hence, using (11), it is just $-\cos\phi /b$.

We are now in a position to write the equation governing
light deflection from spherical gravity waves:
\newline\begin{equation}
u'' + u = {h\over b\sin\phi}
\ \left( 1 - {3\over 2}\sin^{2}\phi - i\Omega b \left[
{1+\cos\phi\over\sin\phi}  - {\sin\phi\over 2} \right] \right)  \  \ .
\end{equation}\newline

This equation can be solved perturbatively by looking
for a function $u_{1}$ satisfying
\newline\begin{equation}
u = u_{0} + u_{1} \  \ ,
\end{equation}\newline
where $u_{0}$ is the background solution, i.e.,
\newline\begin{equation}
u_{0}'' + u_{0} = 0 \  \  ,
\end{equation}\newline
which is the equation of a straight line (see (11)) . Then, the equation for
$u_{1}$ is just (16), with $u$ replaced
by $u_{1}$ on the left-hand side.
\clearpage

\centerline{\bf III. Interpretation}
\vspace*{0.5cm}
The implications of (16) can be seen most clearly in the
optimal case of very close encounter, when the impact
parameter is less than one gravitational wavelength:
$b<\Lambda\equiv 2\pi /\Omega$. (Equation (16) can also
be integrated analytically outside this region [32], but that will
not be considered here.) As we shall see, such favorable
configurations are not all that implausible (see also [28,32].)

For what we called earlier case-one scenarii
(SG and SL unrelated, far-apart objects,) if we
consider a binary star like $\mu$-Sco (see below) as our gravity-wave
source SG, then
we have to find a candidate light source SL within an angle
equal to SG's wavelength (about $18$ light-hours) divided by SG's
distance to the Earth (about $109$ parsecs.) This gives an angular
size of about 1 arcsec. This constraint is less stringent still
if we consider
a case-two situation (SG and SL tightly bound together.) Although
the wavelength here is much shorter (say $0.1$ light-second for
a $10$ Hz neutron star,) the distances are short enough (many such
systems are only about $1$ light-second across,) that the constraint
on the orbit's inclination with respect to the Earth for obtaining
a near-zone deflection is not too extreme (details below.)
Among the over 500 known pulsars, several are known
to be (and statistically many more should be)
eclipsing or near to eclipsing binaries. In the logic of
this paper, these obviously hold
excellent prospects for gravity-wave detection.

Going back to the inspection of (16) in the near-zone,
a particular photon
crosses the region of strongest $h$ in less time
than it takes the gravitational wave profile that he encounters
to change appreciably. (Note that the $1/r$ fall-off of $h$
makes it drop by a few orders of magnitude within a single
wavelength). This means that the photon sees essentially a
standing bulge of curvature (centered on the source), hence (16)
 can be integrated
with $h\approx H/r \ \equiv H\sin\phi /b$ (see(1,11)).
 Subtracting this maximal trajectory
from the one for when $h$ is at its lowest gives the desired shift.

Let us then compute the contribution of each term in (16)
to $u_{1}$, and  subtract the two values of $\phi$ at $u=0$
($r=\infty$) to find the total shift
for near-field deflection. We obtain
\newline\[
\Delta\phi ={H\over b} - {H\over b}
 - i\Omega H ({\pi\over 2}+1) + i\Omega {\pi\over 8}
\]

\begin{equation}
+  {H\over b} - {H\over b}
 - i\Omega H ({\pi\over 2}-1) + i\Omega {\pi\over 8} \  \ ,
\end{equation}\newline
where we have kept the order of the terms in (16).

First, we see that the effects of the $\Omega$-independent terms
in (16) cancel out exactly.  This means that there is no Scwarscild-like
light deflection. (In the Schwarzshild case, the RHS in (16)
reduces to a $u^{2}$ term analogous to our $hu\propto \sin^{2}\phi$.)

Summing up all contributions, we obtain a formula for
maximal shifts from near-field deflections:
\newline\begin{equation}
|\Delta\phi|_{max} \approx {3\over 4}\pi\Omega H \  \ ,
\end{equation}
or
\begin{equation}
|\Delta\phi|_{max} \approx {3\over 2} \pi^{2} h_{max}(r=\Lambda)  \  \ .
\end{equation}\newline
Note that the amplitude does not increase as one penetrates
deeper into the near-zone. This is because, a smaller value
of $b$ corresponding to a smaller angle of incidence (angle
between a light ray and a given spherical wave-front),
the  effect (on $\Delta\phi$) of the corresponding increase in $h$
gets cancelled by the decrease in the sine of the angle
(see \cite{kn:fakir1}.)

As mentioned earlier, choosing a candidate site
involves weighing  the sometimes contradictory effects of
a number of parameters, and finding a compromise
that best accommodates the observational constraints.
One obvious problem is that longer gravitational wavelengths,
 which
correspond  to wider near-zones, and hence to less stringent
constraints on the alignment, often correspond to weaker
gravity-wave amplitudes (see (2)). While we leave for \cite{kn:fakir4}
a more detailed evaluation of various candidates' merits,
we try here to estimate typical orders of magnitude for this
effect in optimal, but realistic situations.

Consider first, as our SG, the binary of giant stars $\mu$-Sco
in the Scorpius constellation. It is situated at a distance from
the Earth $r_{E}=109$ parsecs, and has an expected gravity
wave frequency $f\approx 1.6\times 10^{-6}Hz$. The expected
dimensionless amplitude of gravity waves
at $r=r_{E}$ is $h(r_{E})\approx 2\times 10^{-20}$, implying
that $H\approx 6cm$.
Such a source, which is among the strongest in the Earth's
vicinity, has unfortunately
no chance of being detected by Earth-bound experiments,
due chiefly to seismic noise.

Let us see, in this instance, how far our effect falls
from present limits of resolution power astrometry.
 The best resolution achieved today
comes from Very Long Baseline Interferometry. There, the limit
comes essentially from the shortest wavelengths that survive
water and oxygen absorption (say $0.1$mm,) and the
size of the Earth. The ratio gives about $10^{-5}$ arcsec.
There are however definite projects for reaching $10^{-7}$ arcsec
in a few years with space-based interferometry.

Now, the numbers for $\mu$-Sco imply a near-zone which spreads over
an angular distance of $ c/rf\sim 1$ arcsec, where c is the speed of light.
This does not seem too tiny to allow for a non-negligible chance
of discovering a potential candidate for the role of SL. (Note that
the Hipparcos satellite routinely maps the sky at a resolution of
$10^{-4}$ arcsec.) For obvious reasons, only the binaries that are
closest to the Earth have so far receaved attention in the literature.
But the closeness of SG to the Earth is not a sine qua non in the
present context. Hence, a great many more binaries than the ones
catalogued so far could be observationally interesting here.

The shift's amplitude expected from (20) is
\newline\begin{equation}
|\Delta\phi|_{max}\approx {3\over 2} \pi^{2}H/\Lambda
\approx 2\times 10^{-8}{\rm arcsec} \  \ ,
\end{equation}\newline
which falls short of the limit of detectability by
only one order of magnitude.

The situation for a case-two scenario (SG and SE in a binary) turns out
to be even more promising.
The period and strength of
neutron-stars as gravity-wave sources are poorly known. The limits
on neutron-star parameters are mostly drawn from
pulsar data, but it is still uncertain how a typical ordinary
neutron star compares to a typical pulsar. For instance, neutron-star
rotation frequencies are thought to range all the way up to about
 $10^{3}Hz$,
the  observed higher limit for pulsars, but the lower limit is
very uncertain.)
Also, neutron-star gravity-wave strengths could vary widely
depending on which of the possible
mechanisms of gravity-wave production is actually at work.
Eventually, only observation (why not of gravity waves,) will
put stringent constraints on neutron-star parameters.

To fix ideas, take SG to be a neutron star with
$f\approx 1$Hz, and SL to be a pulsar with a much
higher frequency. SG and SL form a tight binary
system. Their average separation is only about
one light-second.
(The famous binary pulsar PSR 1913+16, e.g., is
less than $3$ light-seconds across, and several tighter
binaries have been observed.)

The alignment constraint here is really  a constraint
on the orbit's inclination
with respect to the Earth. For the light
from SL to cross the near-zone of SG on its way to us, the sine
of that
angle should be smaller than $\Lambda/R$, where $R$ is the orbit's
half-size. But this ratio, given the numbers above,
is typically close to one. Hence, we have here a set of
candidates that naturally satisfy the alignment requirement.

Now, if we assume that the neutron star is generating gravity
waves through the CFS (Chandrasekhar-
Friedman-Schutz) instability [30,31], then a
possible value for $H$ would be $H \approx 10^{-6}m$
(see [4] and references therein.) For $f=1 Hz$, this yields
\newline\begin{equation}
|\Delta\phi|_{max}\approx 10^{-7} {\rm arcsec} \  \  .
\end{equation}\newline

And thus, even neutron-star gravity waves, some of the
faintest to reach the Earth, could be detected afterall,
and relatively soon.

Finally, interesting candidates for the case-two scenario
(SG and SL gravitationally bound) are not limited to tight
binaries of neutron stars. Take the X-ray source Her X-1.
This is thought to consist of a neutron star ($f\approx 1$Hz)
and a $2.2$ solar-mass star, forming a binary system
with an orbital period of $1.7$ day. Because the orbit is
larger here than in the case above (smaller $\Lambda/R$),
strong shifts such as (23) are more likely if the system
is not too far from the eclipsing position. As it turns out,
this is actually an {\em eclipsing} binary, so
excellent alignment occurs naturally every $1.7$ day.
We could also cite, in this class of candidates, several
of the many {\em multiple} stars observed
in the galaxy. These usually contain one or more binaries.
The $\alpha$-Gem system, for instance, is made of
three binary stars, with periods of $0.8$, $2.9$ and $9.2$ days.
Such
configurations as this are, obviously, potential sites for rich
gravity-wave physics in the sense of this paper.

\vspace*{2cm}
\centerline{\bf Acknowledgements}
\vspace*{0.5cm}
The author benefited from instructive discussions with
S. Braham, P. Gregory, H. Richer and W. Shuter.
The author is particularly grateful
to W.G. Unruh for showing interest in this modest work,
and providing valuable advice. This work would not have
been possible without his continuous support.

This research was supported by the Cosmology Group in the
Department of Physics, University of British Columbia, and
by the Moroccan Ministry of Education, Rabat, Morocco.


\clearpage
\begin{thebibliography}{99}
\vspace*{1cm}
\bibitem{kn:fakir0} R. Fakir (1991), {\em Gravity Waves and Hipparcos},
proposal presented in {\em Table Ronde on Moroccan-Europeen Collaboration
in Astronomy, Moroccan Center for Scientific Research Archives, June 1991,
Rabat, Morocco.}
\bibitem{kn:fakir1} R. Fakir (1992), {\em Gravitational Wave Detection,
a Non-Mechanical Effect}, {\em The Astrophysical Journal},
vol.418, 20.
\bibitem{kn:brgr} V.B. Braginski and L.P. Grishchuk,
{\em Soviet Physics-JETP} 62, 427 (1985).
\bibitem{kn:th300} K.S. Thorne, in {\em 300 Years of Gravitation}, S.W.
Hawking and W. Israel, Eds.(Cambridge University Press, Cambridge, 1987.)
\bibitem{kn:thsci} K.S. Thorne, {\em Science} 256, 325 (1992).
\bibitem{kn:thbb} K.S. Thorne, in {\em Recent Advances in General Relativity},
A. Janis and J. Porter, Eds.(Birkhauser, Boston, 1992.)
\bibitem{kn:grbb} L.P. Grishchuk, {Annals of the New York Academy of
Sciences}, vol.302, 439 (1977).
\bibitem{kn:ei1} A. Einstein,
{\em Preuss. Akad. Wiss. Berlin, Sitzungsberichte der
physicalisch-mathematischen Klasse}, p.688 (1916).
\bibitem{kn:ei2} A. Einstein,
{\em Preuss. Akad. Wiss. Berlin, Sitzungsberichte der
physicalisch-mathematischen Klasse}, p.154 (1918).
\bibitem{kn:we} H. Weyl, {\em Space-Time-Matter}. Methuen:London (1922).
\bibitem{kn:ed} A.S. Eddington, {\em The Mathematical Theory of Relativity},
2nd edn, Cambridge University Press, Cambridge (1924).
\bibitem{kn:mtw} C.W. Misner, K.S. Thorne, J.A. Wheeler, {\em Gravitation}
(Freeman, San Fransisco,1973.)
\bibitem{kn:thzim} M. Zimmerman and K.S. Thorne, in {\em Essays in General
Relativity}, F.G. Tipler, ed. (Academic Press: New York,1980.)
\bibitem{kn:bon} J.R. Bond and B.J. Carr, {\em Monthly Notices of the
Royal Astronomical Society} 207, 585 (1984).
\bibitem{kn:sma} L. Smarr, {\em Sources of Gravitational Waves}
(Cambridge University Press, Cambridge, 1979.)
\bibitem{kn:derpir} see reviews in N. Deruelle and T. Piran,
{\em Gravitational Radiation} (North Holland: Amsterdam, 1983.)
\bibitem{kn:web} J. Weber, {\em Physical Review} 117, 306 (1960).
\bibitem{kn:detec} see for example J.A. Tyson and R.P. Giffard,
{\em Annual Reviews of Astronomy and Astrophysics} 16, 521 (1978),
E. Amaldi and G. Pizella, in {\em Relativity, Quanta and Cosmology
in the Development of the Scientific Thought of Einstein}, vol.1, 96 (1979),
and references in D.G. Blair, {\em The Detection of Gravitational Waves}
(Cambridge University Press, Cambridge, 1991) and P.F. Michelson,
J.C. Price and R.C. Taber, {\em Science} 237, 150 (1987).
\bibitem{kn:chris} D. Christodoulou, {\em Physical Review Letters} 67,
1486 (1991).
\bibitem{kn:thchr} K.S. Thorne, {\em Physical Review D} 45, 520 (1992).
\bibitem{kn:ligo} R.E. Vogt, {\em The U.S. LIGO Project} in
{\em Proc. of the Sixth Marcel Grossmann Meeting on GRG, MGC, Kyoto,
Japan, 1991.}
\bibitem{kn:virgo} C. Bradaschia et. al., {\em Nucl. Instrum. \& Methods},
A289, 518 (1990).
\bibitem{kn:koth} S.J. Kovacs and K.S. Thorne,
{\em The Astrophysical Journal} 224, 62 (1978).
\bibitem{kn:grpo} L.P. Grishchuk and A.G. Polnarev,
{\em Sov.Phys. JETP} 69, October 1989.
\bibitem{kn:bopr} R.J. Bontz and R.H. Price,
{\em The Astrophysical Journal} 228, 560 (1979).
\bibitem{kn:sm} L. Smarr, {\em Physical Review D} 15, 2069 (1977).
\bibitem{kn:brth} V.B. Braginsky and K.S. Thorne,
{\em Nature} 327, 123 (1987).
\bibitem{kn:fakir4} R. Fakir (1993), {\em Gravity Waves and Light},
in preparation.
\bibitem{kn:fakir3} R. Fakir (1993), {\em Almost Readily Detectable
Time Delays from Gravity Waves?}, UBC preprint UBCTP-93-006.
\bibitem{kn:chandra} S. Chandrasekhar , {\em Physical Review Letters},
{\bf 24}, 611 (1970).
\bibitem{kn:FS} J.L. Friedman and B.F. Schutz,
{The Astrophysical Journal}, {\bf 222}, 281 (1978).
\bibitem{kn:fakir5} R. Fakir (1993), {\em Early Direct Detection
of Gravity Waves}, UBC preprint UBCTP-93-016.

\end{thebibliography}
\end{document}